\documentclass[letterpaper,prc,twocolumn,showpacs,floatfix,nofootinbib,preprintnumbers,superscriptaddress,amsmath,amssymb]{revtex4-1} 
\usepackage{graphicx}
\usepackage{bm}
\usepackage{wrapfig}
\usepackage{amsmath,amssymb}
\usepackage{color}
\usepackage{ulem}
\def\vect#1{{\mbox{\boldmath $#1$}}}

\begin{document}


\title{Global analysis of the non-uniformity of nucleon density distributions
}


\author{Shuichiro~Ebata}
\email[]{ebata@mail.saitama-u.ac.jp}
\affiliation{Department of Physics, Graduate School of Science and Engineering, Saitama University, Saitama, 338-8570, Japan}
\author{Wataru~Horiuchi}
\email[]{whoriuchi@omu.ac.jp}
\affiliation{Department of Physics, Osaka Metropolitan University, Osaka 558-8585, Japan}
\affiliation{Nambu Yoichiro Institute of Theoretical and Experimental Physics (NITEP), Osaka Metropolitan University, Osaka 558-8585, Japan}
\affiliation{RIKEN Nishina Center, Wako 351-0198, Japan}
\affiliation{Department of Physics,
  Hokkaido University, Sapporo 060-0810, Japan}


\date{\today}

\begin{abstract}
  \noindent{\bf Background:}
  Saturation of nuclear density is a fundamental property of atomic nuclei
  but in reality, the nuclear internal density distribution is not uniform, e.g., some nuclei are known to have the so-called bubble structure, in which the central density is depressed.\\
\noindent{\bf Purpose:} We aim to unveil
the emergent mechanism of the non-uniformity of the nucleon density distributions for whole nuclear mass regions, 
not only for a typical bubble structure.\\
\noindent{\bf Method:} We systematically investigate the nucleon density distributions
using the Skyrme Hartree-Fock plus Bardeen-Cooper-Schrieffer calculation
represented in the three-dimensional Cartesian coordinate space.
The ground states of 1,389 even-even nuclei are generated. 
To quantify the non-uniformity of these density distributions, a ``generalized bubble parameter'' is introduced.\\
\noindent{\bf Results:}
We find that the bubble structure appears around the magic numbers, 
which correspond to the regions where the $s$ orbit appears near the Fermi surface.
The nuclear deformation and pairing correlations strongly affect the occupation probability, 
but the robust bubble structure of a medium mass nucleus, $^{100}$Sn, is found.
We confirm that the Coulomb force enhances the bubble degree in the superheavy region. 
The nuclear non-uniformity is further generalized by the ``multi-layered'' bubble structure, 
which exhibits some density depression in the internal regions of the density distributions.\\
\noindent{\bf Conclusion:}
The non-uniformity of the internal density distribution
occurs due to the deficiency of the specific single-particle orbits: the nodal $s$, $p$, and $d$ orbits.
This is certainly reflected in the density distribution near the nuclear surface,
which can be deduced from proton-elastic scattering.
\end{abstract}

\pacs{}

\maketitle

\section{Introduction}
\label{sec:introduction}
The saturation of nucleon density is a fundamental property of
an atomic nucleus, which leads to the empirical nuclear radius
formula, $\propto A^{1/3}$, with $A$ being the atomic mass number.
In 1946, Wilson pointed out
the non-uniformity of the internal density
due to the electrostatic forces~\cite{Wil46}.
This conjecture was realized by microscopic mean-field calculations:
The single-particle orbits are pushed out by a strong Coulomb field
in heavy mass nuclei~\cite{PhysRevC.98.054319,PhysRevC.106.024321}. 
This remarkable characteristic is called the ``bubble'' structure, 
which exhibits a central depression of the internal density.
  The emergent mechanism of the bubble structure
has attracted many researchers as it may be related to 
the nuclear saturation property
along with the evolution of the shell structure.

In the 1970s, the bubble structure for both light and heavy nuclei
was predicted by mean-field models~\cite{DAVIES1972455,CAMPI1973291}.
Subsequent theoretical research also predicted 
many bubble nuclei in medium-mass~\cite{KHAN200837,Gra09},
superheavy~\cite{Ben99}, and hyperheavy regions~\cite{DECHARGE200355}. 
The emergence of the bubble structure is due to the lack of the occupation
of the $s$ orbit.
Possible mechanism of the change of the $s$ orbit occupation
has been discussed in terms of 
nuclear deformation, pairing correlations, as well as
the temperature dependence~\cite{SAXENA20191,SAXENA2019323}.
Furthermore, the correlation between the degree of the bubble
structure and the equation of state (EoS) of nuclear matter
was discussed towards constraining the EoS parameters 
~\cite{Sch17}.
The role of the $s$ orbits for central nuclear density has been
investigated experimentally from the comparison
between the charge density distributions of $^{206}$Pb and $^{205}$Tl~\cite{Eut76}.
For light nuclei, many theoretical studies, including the shell-model calculation,
predicted a significant depletion of the internal density
in $^{34}$Si~\cite{Gra09,Ben99,Yao12,Li16,Dug17}.
Evidence supporting the proton-bubble structure
in $^{34}$Si was reported experimentally, indicating
smaller occupation probability of the $1s$ orbit
than that in $^{36}$S~\cite{Mutschler2017}.

A systematic investigation of the nuclear spectroscopic properties is
necessary to identify the bubble nuclei on the nuclear chart.
For this purpose, it is quite useful to investigate
the nuclear density distribution near the nuclear surface.
It was shown that the nuclear surface diffuseness
  defined in Ref.~\cite{PhysRevC.97.054607} is
closely related to the nuclear spectroscopic properties
near the Fermi level~\cite{ptab136, PhysRevC.104.054313}
and allows one to discuss the bubble structure~\cite{PhysRevC.102.034619}.

In this paper, we systematically study the density distributions
of 1,389 kinds of even-even nuclei to 
clarify the emergent mechanism of the bubble structure in the different mass regions.
In addition to the discussions on the ordinary bubble structure,
we introduce a new point of view, the ``multi-layered" bubble (MLB) structure, and study it,
which is regarded as a higher-order bubble structure.
  With this generalized bubble structure, the non-uniformity
  of the nuclear density distributions 
  is quantified for the whole mass region.

The paper is organized as follows.
Section~\ref{sec:2} summarizes the formulations necessary
to investigate nuclear density profiles.
The mean-field model employed in this paper
is briefly explained in Sec.~\ref{sec:2.1}.
The numerical setup for about 1,400 even-even nuclei is
shown in Sec.~\ref{sec:2.3}.
The quantifications of the internal
  density depression and the nuclear surface diffuseness  
are given in Sec.~\ref{sec:2.2}. 
The calculated internal density depletion is 
presented in Sec.~\ref{sec:results}. 
The emergence of bubble nuclei on the nuclear chart is overviewed in Sec.~\ref{sec:bub}
and details of their structure from light to heavy nuclei are discussed in Sec.~\ref{Dbe}.
Section~\ref{sec:hollow} discusses
the MLB structure is a natural extension of
the bubble structure.
Finally, the summary and conclusion are made in Sec.~\ref{sec:concl}.

\section{Formulation}
\label{sec:2}
\subsection{Mean-field model}
\label{sec:2.1}
To obtain the nuclear density distributions for a wide range of mass regions,
we employ Hartree-Fock plus Bardeen-Cooper-Schrieffer theory (HF+BCS)
with Skyrme energy density functional.
To describe various nuclear shapes, the wave function is expressed in the three-dimensional (3D)
Cartesian coordinate space $(x,y,z)$~\cite{Ebata2017}.
The space is discretized in a cubic mesh $\Delta x=\Delta y=\Delta z$.
The $i$th single-particle wave function is represented as 
$\langle x,y,z; \sigma| \phi_{i,\tau} \rangle = \phi_{i,\tau}(x,y,z,\sigma)$,
where $\sigma$ and $\tau$ denote the spin ($\uparrow$ and $\downarrow$) and 
isospin (neutron: $n$ and proton: $p$) coordinates, respectively.
The intrinsic nuclear density at $\vect{r} = (x,y,z)$ is calculated as
\begin{eqnarray}
\label{eq:rho}\rho^\tau(\vect{r}) \equiv \sum_{i,\sigma} v_{i,\tau}^2 |\phi_{i,\tau}(x,y,z,\sigma)|^2, 
\end{eqnarray}
where $v_{i,\tau}^2$ is the occupation probability of the $i$th state and for each $\tau$.
The matter density distribution
$\rho^m(\vect{r})$ is a sum of $\rho^n(\vect{r})$ and $\rho^p(\vect{r})$.
The intrinsic density is calculated fully self-consistently according to HF and BCS gap equations~\cite{RS80}
\begin{eqnarray}
&&\label{eq:hf}[ h,\rho ] = 0,\\
&&\label{eq:gap}2\varepsilon_{i,\tau} u_{i,\tau} v_{i,\tau}+\Delta_{i,\tau}(2v_{i,\tau}^2-1) =0,\ \ \ i>0,
\end{eqnarray}
where $h$, $\rho$ are the single-particle Hamiltonian and normal density matrix, respectively. 
$\varepsilon_{i,\tau}$ is the single-particle energy of the $i$th state, 
which is equal to the value of the $i$th pair state in even-even nuclei.
$\Delta_{i,\tau}$ is a pairing gap parameter. 
The procedure of the computations is given as follows:
I) solve HF equation of Eq.~(\ref{eq:hf}) with the normalization condition, 
II) update the HF single-particle Hamiltonian with the density Eq.~(\ref{eq:rho})
and calculate a single-particle energy of $i$th state:
$\varepsilon_{i,\tau} = \langle \phi_{i,\tau} | h | \phi_{i,\tau} \rangle$, 
III) solve the BCS gap equation of Eq.~(\ref{eq:gap})
with the single-particle energy and its pairing gap, 
IV) back to I) with the updated occupation probabilities.
This iteration procedure is repeated until the total energy is converged~\cite{Ebata2017,RS80}.

\subsection{Numerical setup}
\label{sec:2.3}

We investigate the nuclear densities of even-even nuclides with $Z=6$--118.
For $Z<50$, we calculate the isotopes whose neutron or proton chemical potential 
is higher than 2 MeV to avoid unphysical nucleon gas problems in the HF+BCS 
applications~\cite{10.3389/fphy.2020.00102}.
For $Z\geq 50$, the isotopes up to $N=2Z$ are computed. 
The total number of the calculated even-even nuclides is 1,389.

In the numerical computations, the mesh size $\Delta x$ and box size $R_{\rm cal}$ are 
are appropriately for the size of nucleus chosen as:
$\Delta x=0.8$ fm with $R_{\rm cal}$ = 12 fm for $Z=6$--18,  
$\Delta x=1.0$ fm with $R_{\rm cal}$ = 15 fm for $Z=20$--80, and
$\Delta x=1.0$ fm with $R_{\rm cal}$ = 20 fm for $Z=82$--118.
Although the radius of $^{354}$Og ($Z$=118) is loughly estimated to be 8.5 fm, 
the computational space we adopted is sufficiently large even when nuclear deformation is taken into account.
The SkM$^\ast$~\cite{BARTEL198279} 
parameter set is used as a primary effective interaction otherwise mentioned.

\subsection{Nuclear density and generalized bubble parameter}
\label{sec:2.2}
Once the converged intrinsic density is obtained, 
we convert it to the laboratory frame by taking an average over angles as~\cite{PhysRevC.86.024614}
\begin{eqnarray}
\rho^\tau(r) \equiv \frac{1}{4\pi} \int\! d\hat{\vect{r}}\ \rho^\tau(\vect{r}).
\end{eqnarray}

\begin{figure}[h]
\begin{center}
\includegraphics[height=0.45\textwidth,keepaspectratio,angle=-90]{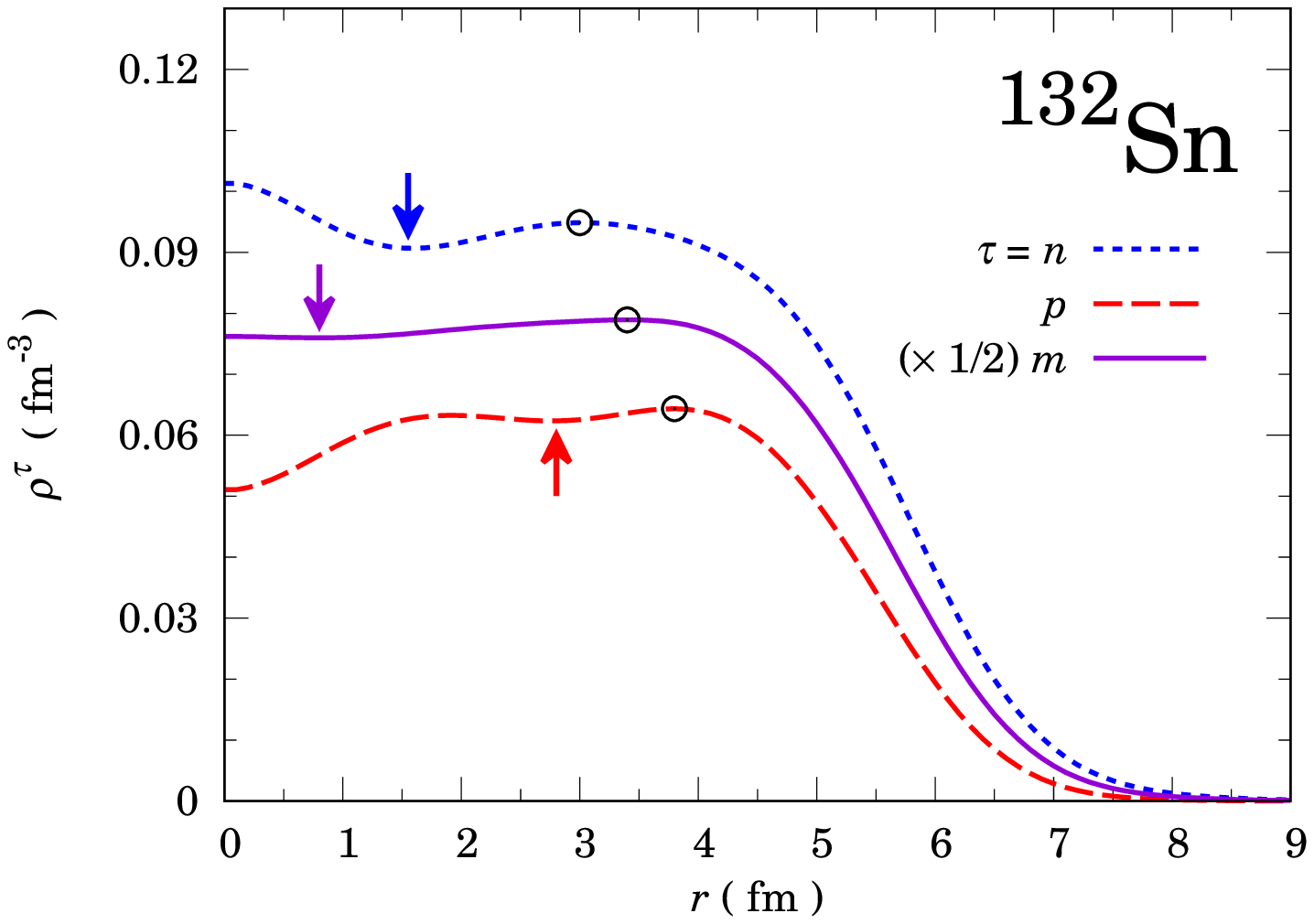}
\caption{Nucleon density distributions for $^{132}$Sn with SkM$^\ast$ parameter set.
The matter density distribution is halved for visibility. 
The arrows indicate the local minima of the density distributions.
The $\varrho^\tau_{\rm max}$ is indicated by an open circle for each distribution.  
}
\label{DSn132}
\end{center}
\end{figure}

For the sake of convenience, we introduce a ``generalized bubble parameter'', $G(r^\tau_{\rm min})$, 
to quantify the non-uniformity of the nuclear density distributions
including the bubble structure 
\begin{equation}
G^\tau(r^\tau_{\rm min}) \equiv \frac{\varrho^\tau_{\rm max}-\rho^\tau(r^\tau_{\rm min})}{\varrho^\tau_{\rm max}},
\label{eq:Gb}
\end{equation}
where $\varrho^\tau_{\rm max}$ is the maximum density
in the range from $r^\tau_{\rm min}$ to the root-mean-square radius of the density distribution and
$r^\tau_{\rm min}$ is defined as a radial distance at local minima. 
In general, a nucleus has multiple $r^\tau_{\rm min}$ 
in which density depression may occur not only at the center of a nucleus but also in the middle.
When $r^\tau_{\rm min}=0$ and $\rho^\tau(r^\tau_{\rm min}) < \varrho_{\rm max}^\tau$,
it is nothing but the quantity defined in Ref.~\cite{SAXENA20191}
and quantifies a central density depression, namely, the bubble structure,
Hereafter, $G^\tau(0)$ is simply denoted as $G^\tau_{\rm B}$.
Figure~\ref{DSn132} plots the matter, neutron, and proton density distributions of $^{132}$Sn.
The $r^\tau_{\rm min}\neq0$ and $\varrho^\tau_{\rm max}$ values 
are indicated by arrows and open circles in Fig.~\ref{DSn132}.
The proton density distribution exhibits the central depression, 
and its bubble parameter is evaluated as $G^p_{\rm B}=0.21$, $\varrho^p_{\rm max}=\rho^p(r=3.8\ {\rm fm})$. 
But for neutron, the $G^n_{\rm B}$ is zero by definition, 
as the density distribution does not show the central depression.
In total, the $G^m_{\rm B}$ is finite, although it is small, less than 0.05.
In this paper, we call this sort of nucleus a semi-bubble nucleus, 
which only has a bubble structure for neutrons or protons.

As shown in Fig.~\ref{DSn132}, the density distributions of $^{132}$Sn exhibit some local minima. 
This MLB structure is indicated by an arrow for each distribution of Fig.~\ref{DSn132}.
The number of the local minima excluding at $r^\tau_{\rm min}=0$ 
is called a bubble multiplicity $M^\tau_{\rm B}$, which is used to classify the nuclear 
bubblicity in the present paper.
When $M^\tau_{\rm B}>1$, the nucleus has the MLB structure
as indicated in the all density distributions of $^{132}$Sn ($M^{n,p,m}_{\rm B}=1)$
in Fig.~\ref{DSn132}, although the neutron density has no bubble at the central part.

When we discuss the MLB structure, the generalized bubble parameter
is denoted as $G^\tau_{\rm MLB}$ and its value is calculated with 
the $r^\tau_{\rm min}$, giving the minimum density between its center and $\varrho_{\rm max}^\tau$, 
excluding the value at the origin ($r_{\rm min}^\tau=0$).
For $^{132}$Sn, the $G^p_{\rm MLB}=0.031$ and $G^n_{\rm MLB}=0.044$ are calculated 
using the densities at $r_{\rm min}^p=2.8$ and $r_{\rm min}^n=1.6$ fm. 

The bubble structure is reflected in the density profile near the nuclear surface~\cite{PhysRevC.102.034619}.
To quantify it, the nuclear diffuseness parameter is deduced by introducing the two-parameter Fermi function, 
\begin{eqnarray}
\tilde{\rho}^\tau(r) = \frac{\tilde{\rho}_0^\tau}{1+e^{(r-\tilde{R}^\tau)/a^\tau}},
\label{eq:diff}
\end{eqnarray}
where $\tilde{\rho}_0^\tau$ is determined by the normalization condition for given $a^\tau$ and $\tilde{R}^\tau$.
The diffuseness $a^\tau$ and radius $\tilde{R}^\tau$ parameters are fixed by the fitting procedure
prescribed in Ref.~\cite{PhysRevC.97.054607} using $\rho^\tau(r)$.

\section{Results and discussions}
\label{sec:results}
First, the bubble parameters $G_{\rm B}$ are investigated to provide an overview 
of the emergence of the bubble structure on the nuclear chart and then the nuclear density distributions 
for specific isotopes and isotones are examined to see the emergent mechanism of the nuclear bubble structure.
The effects of nuclear deformation, pairing correlations, and Coulomb force on the bubble formation are investigated.
Finally, the nuclear MLB structure is discussed. 
\subsection{Bubble structure on the nuclear chart}
\label{sec:bub}
\begin{figure}[h]
\begin{center}
\includegraphics[height=0.5\textwidth,keepaspectratio,angle=-90]{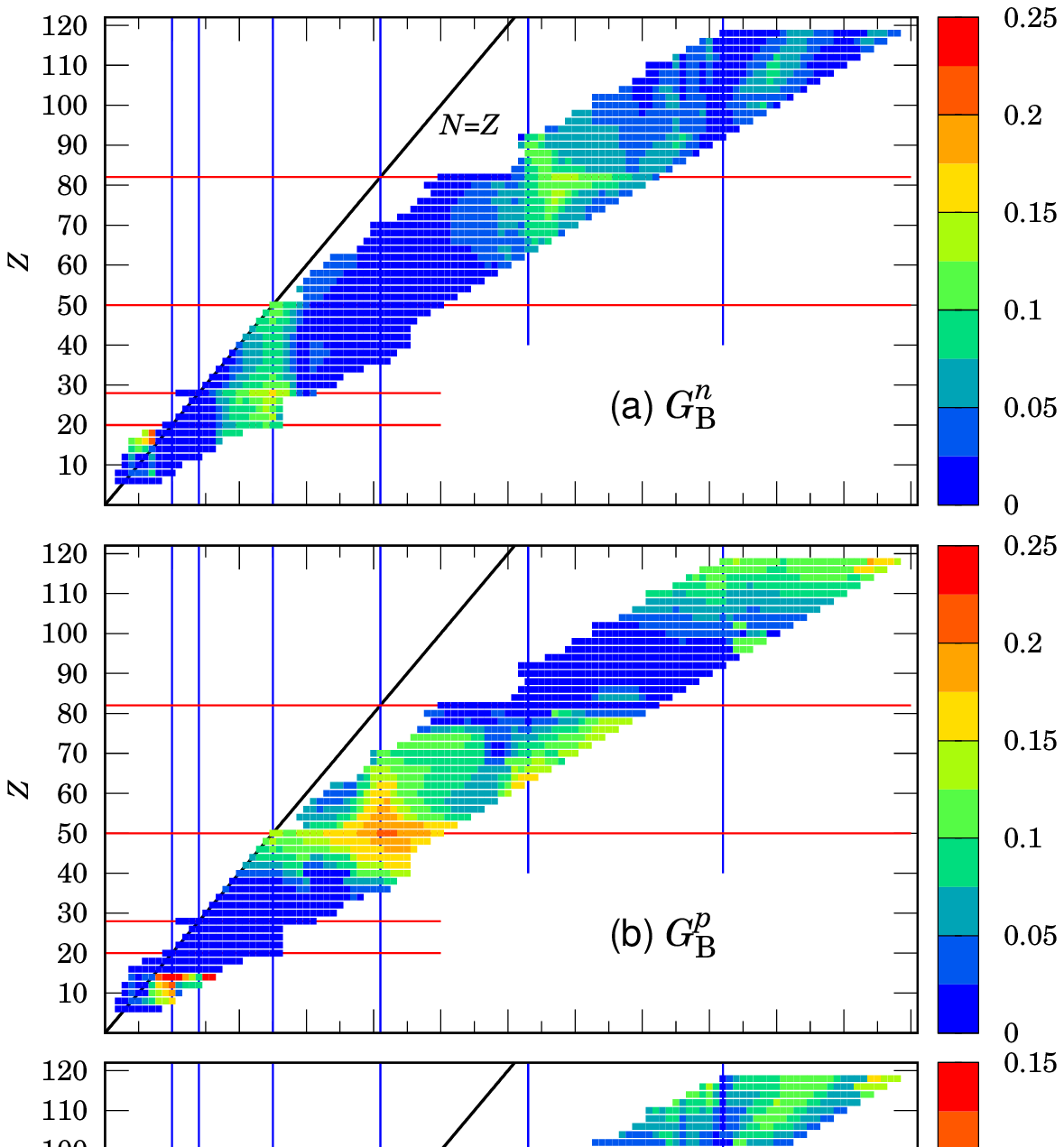}
\caption{Bubble parameters for (a) neutron $G_{\rm B}^n$, (b) proton $G_{\rm B}^p$, 
and (c) matter $G_{\rm B}^m$ density distributions
on the nuclear chart.  
The vertical and horizontal lines denote the neutron and proton magic numbers
(20, 28, 50, 82, 126, and 184), respectively. 
}
\label{bubblechart}
\end{center}
\end{figure}

Figure~\ref{bubblechart} plots the bubble parameters $G^\tau_{\rm B}$ 
for (a) neutron, (b) proton, and (c) matter density distributions.
The emergence of the bubble structure is closely related to the magicity of neutron and proton numbers
as the bubble structure prefers a spherical shape:
The large $G^n_{\rm B}$ ($G^p_{\rm B}$) values are found at around $N=14,16,50$, and $126$ ($Z=14,16$, and $50$) regions.
The proton bubble structure also appears in superheavy ($Z>100$) regions.
We see large $G^m_{\rm B}$ values roughly at four regions;
(i) at $N=14,16$ and $Z=14,16$, (ii) at around Sn isotopes,
(iii) $Z<82, N>100$, and (iv) superheavy regions.
The large $G^m_{\rm B}$ areas can be roughly explained by the combination of 
of $G^n_{\rm B}$ and $G^p_{\rm B}$ patterns.
The maximum amplitude of $G^m_{\rm B}$ appears at $^{100}$Sn, while $G^m_{\rm B}$ of $^{132}$Sn gets small 
as the neutron distribution fills the depressed proton distribution in the internal regions.

\begin{figure}[h]
\begin{center}
\includegraphics[height=0.5\textwidth,keepaspectratio,angle=-90]{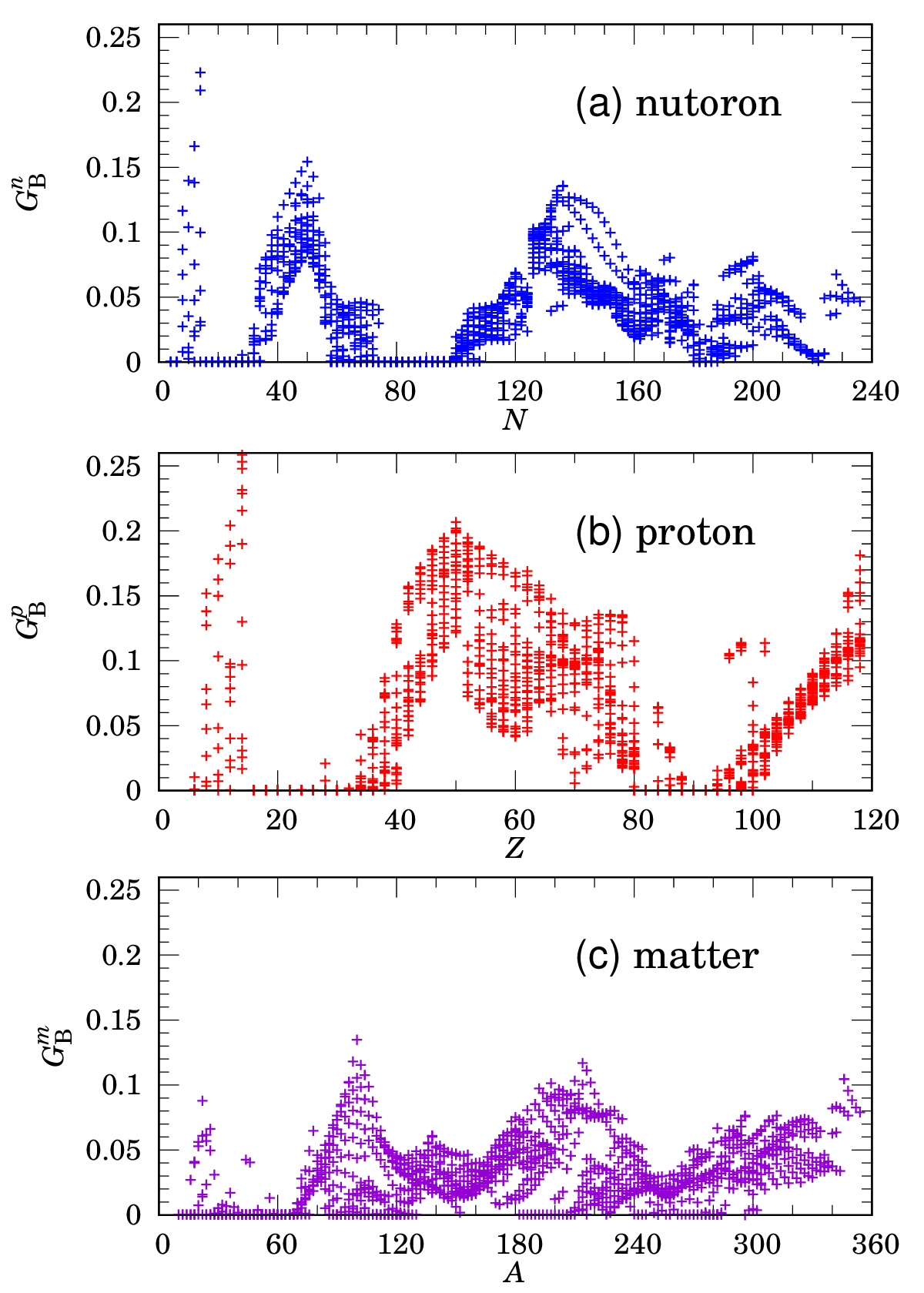}
\caption{Bubble parameters for (a) neutron, (b) proton, and (c) matter density distributions
  as a function of neutron, proton, and mass numbers, respectively. 
}
\label{bubble-anz}
\end{center}
\end{figure}
To see them more quantitatively, in Fig.~\ref{bubble-anz} we also plot 
$G_{\rm B}^\tau$ as a function of the mass number.
The $G_{\rm B}=0$ regions $N\approx 4$--$6$, 16--30, and 74--98 in panels (a) and (b)
respectively correspond to the regions where the $1s$, $2s$, and $3s$ orbits are occupied.
This anti-bubble effect was discussed in many previous studies~\cite{PhysRevC.98.054319,PhysRevC.106.024321,DAVIES1972455,CAMPI1973291,KHAN200837,Gra09,Ben99,DECHARGE200355,SAXENA20191,SAXENA2019323}.
We see several peak structures between the $G^{\tau}_{\rm B}=0$ regions.
$G^p_{\rm B}$ is more pronounced than $G^n_{\rm B}$ because protons are more deeply bound than neutrons.
We will discuss the emergent mechanism of the bubble structure
for specific nuclei where the bubble structure is pronounced.
  
\subsection{Evolution of the bubble structure}
\label{Dbe}
The occupation of the $s$ orbits is an essential viewpoint 
to understand the emergent mechanism of the bubble structure~\cite{PhysRevC.98.054319,PhysRevC.106.024321,DAVIES1972455,CAMPI1973291,KHAN200837,Gra09,Ben99,DECHARGE200355,SAXENA20191,SAXENA2019323}.
In this subsection, the evolution of the nuclear bubble structure along isotope
and isotone chains are discussed in detail. 

\subsubsection{Bubble structure in the region (i): $N,Z=14,16$}
\label{bbldetail}

\begin{figure*}[ht]
\begin{center}
  \includegraphics[height=0.9\textwidth,keepaspectratio,angle=-90]{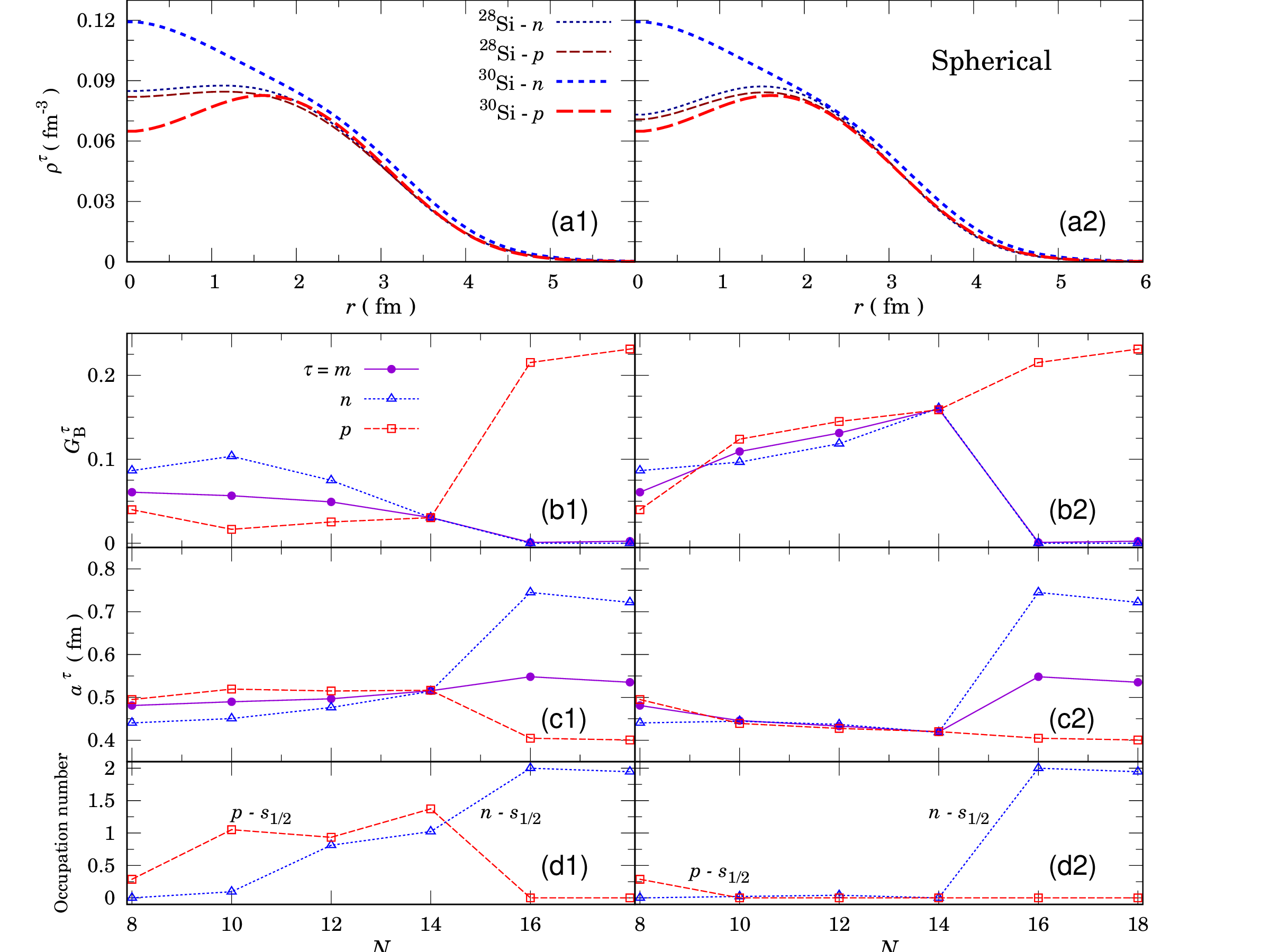}
  \caption{(a) Density distributions, (b) bubble and (c) diffuseness parameters, 
  and (d) occupation numbers of $s$-states obtained from (a1)--(d1) HF+BCS 
  and (a2)--(d2) spherical constrained HF+BCS calculations. 
  Connecting lines for (b), (c), and (d) are plotted as a guide to the eye.
}
\label{BDnz14}
\end{center}
\end{figure*}

Here, we investigate the nuclear density profiles for Si ($Z=14$) isotopes. 
Figure~\ref{BDnz14} displays the density profiles, bubble and diffuseness parameters, 
and occupation numbers of $s$-states for Si isotopes. 
Panels (a1)--(d1) show HF+BCS results,
while the results obtained with the spherical constraint are displayed in panels (a2)--(d2).
Details for the spherical constraint calculations are given in Ref.~\cite{Ebata2017}.
We see that the bubble parameters of the spherical constrained HF+BCS calculations tend to 
be larger than those of deformed ones ($N=10, 12$ and 14).
As pointed out in Ref.~\cite{PhysRevC.102.034619}, 
the larger bubble parameter, the smaller diffuseness parameter becomes.
We calculate the occupation number of the specific spherical single-particle
orbit for the deformed states, which is extracted as
\begin{eqnarray}
\eta_{j}^\tau \equiv (2j+1) \sum_{i} v_{i,\tau}^2
\langle \phi_{j,\tau}^{\rm R}(\vect{r}) | \phi_{i,\tau} (\vect{r})\rangle, 
\label{eq:eta}
\end{eqnarray}
where $|\phi_{i,\tau} \rangle$ is the single-particle states of the HF+BCS state,
and $|\phi_{j,\tau}^{\rm R}\rangle$ is a reference orbit with a spin $j$ 
which is a state in the spherical constrained HF calculation. 
The occupation probability of the reference state $v_{j,\tau}^2$ 
is replaced to unity in Eq.(\ref{eq:eta}).

We obtain the deformed ground states for 
$^{24}$Si, $^{26}$Si, and $^{28}$Si. 
For instance, the reference orbit is chosen as a $2s$ state of $j=1/2$, 
and they have $\eta_{2s}^n$ are 0.096, 0.812, and 1.022, and also, 
the $\eta_{2s}^p$ are 1.052, 0.934, and 1.374, respectively.
The occupation of the $2s$ orbit reduces the bubble parameter and enhances the diffuseness parameters 
since no centrifugal barrier exists near the nuclear surface~\cite{ptab136}.
This behavior is seen in Fig.~\ref{BDnz14} (a1)--(d1). 
When $^{24}$Si, $^{26}$Si, and $^{28}$Si have a spherical shape, 
the occupation numbers of $s$-states for neutron and proton are reduced, 
while their bubble parameters are larger and diffuseness parameters get smaller. 
As the Coulomb interaction is not large in such light nuclei, 
we see the isospin symmetric behavior on the nuclear density for $N=14$ isotones.

\begin{figure}[h]
\begin{center}
\includegraphics[height=0.5\textwidth,keepaspectratio,angle=-90]{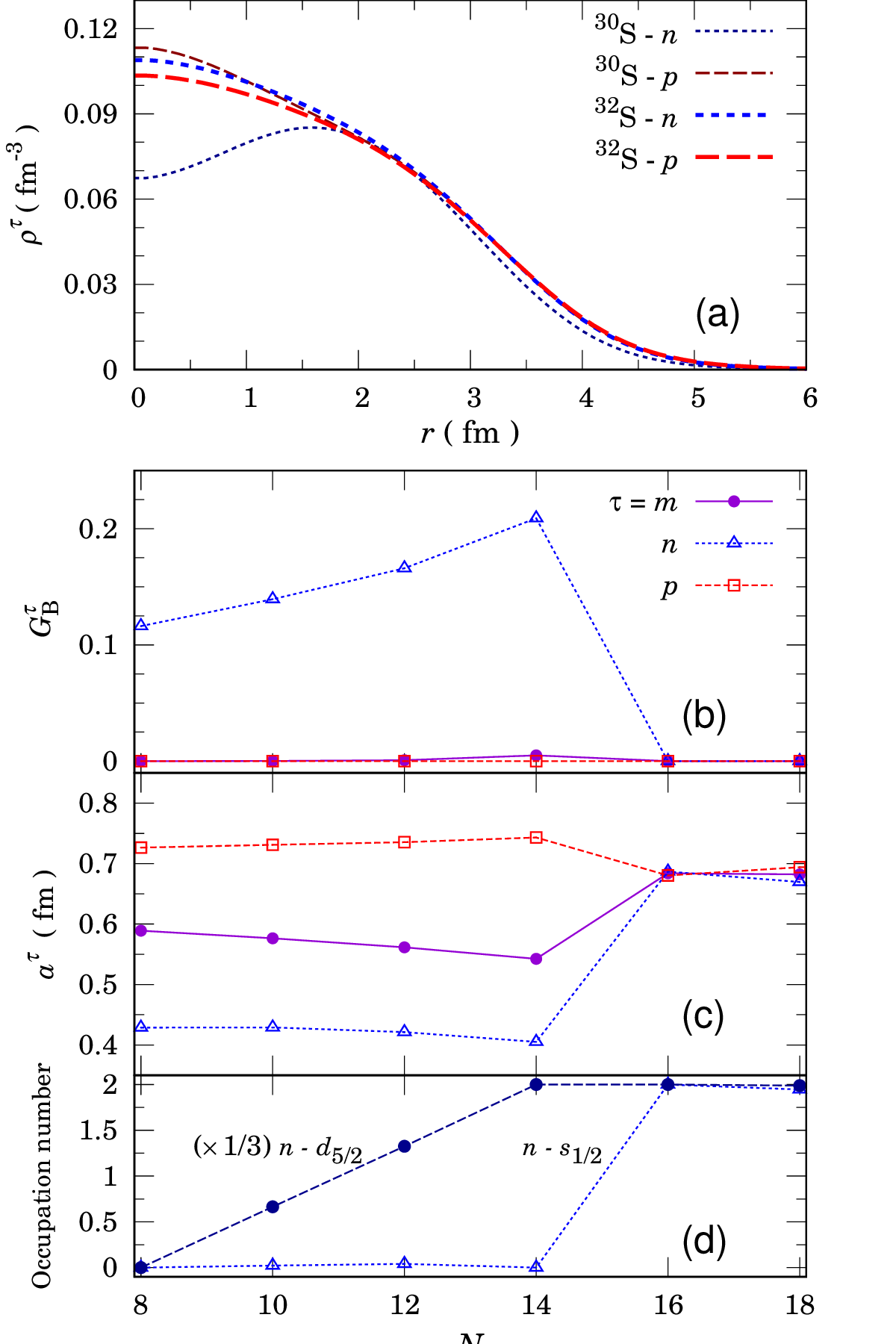}
\caption{Same as Fig.~\ref{BDnz14}, but for $Z=16$ isotopes.
Panel (d) shows occupation numbers of $1d_{5/2}$ and $2s_{1/2}$ states for neutron.
The numbers are obtained by Eq.(\ref{eq:eta}).
}
\label{DBDnz16}
\end{center}
\end{figure}

Next, we focus on the S ($Z=16$) isotopes, and 
their nuclear density properties are drawn respectively on the panels (a)--(d) of Fig.~\ref{DBDnz16}. 
In the present calculation, all the S isotopes are found to be spherical.
The emergent mechanism of the bubble structure is simpler than that of the Si one.
Figure~\ref{DBDnz16} shows the density profiles, bubble and diffuseness parameters, 
and the occupation numbers of $1d$ and $2s$ neutron states for $Z=16$ isotopes.
We see that they have a neutron semi-bubble structure for $N=8$--$14$.
The $G^n_{\rm B}$ becomes maximum at $N=14$, $^{30}$S, 
because the maximum neutron density $\varrho_{\rm max}^n$ increases due to the occupation of the $d$ orbits, 
and disappears at $N=16$, $^{32}$S, since the $2s$ orbits are occupied.
Similarly to the Si isotopes, the Coulomb interaction is not large in such light nuclei. 
We also see the isospin symmetric behaviors on the nuclear densities between these mirror nuclei.

It should be noted that nuclear deformation does not 
always necessarily reduce the degree of the bubble structure.
That depends on the energy position of the occupied and unoccupied $s$ orbits. 
This relation between nuclear deformation and diffuseness was discussed in Ref.~\cite{HI21},
e.g., the nuclear deformation may reduce the bubble formation for Si isotopes.
In contrast, the occupation probabilities of the $s$ state can be reduced by 
the nuclear deformation for S isotopes, where the proton $2s$ orbits are already occupied in the spherical limit.

\subsubsection{Bubble structure in the region (ii): $Z=50$ isotopes}

\begin{figure}[h]
\begin{center}
\includegraphics[height=0.5\textwidth,keepaspectratio,angle=-90]{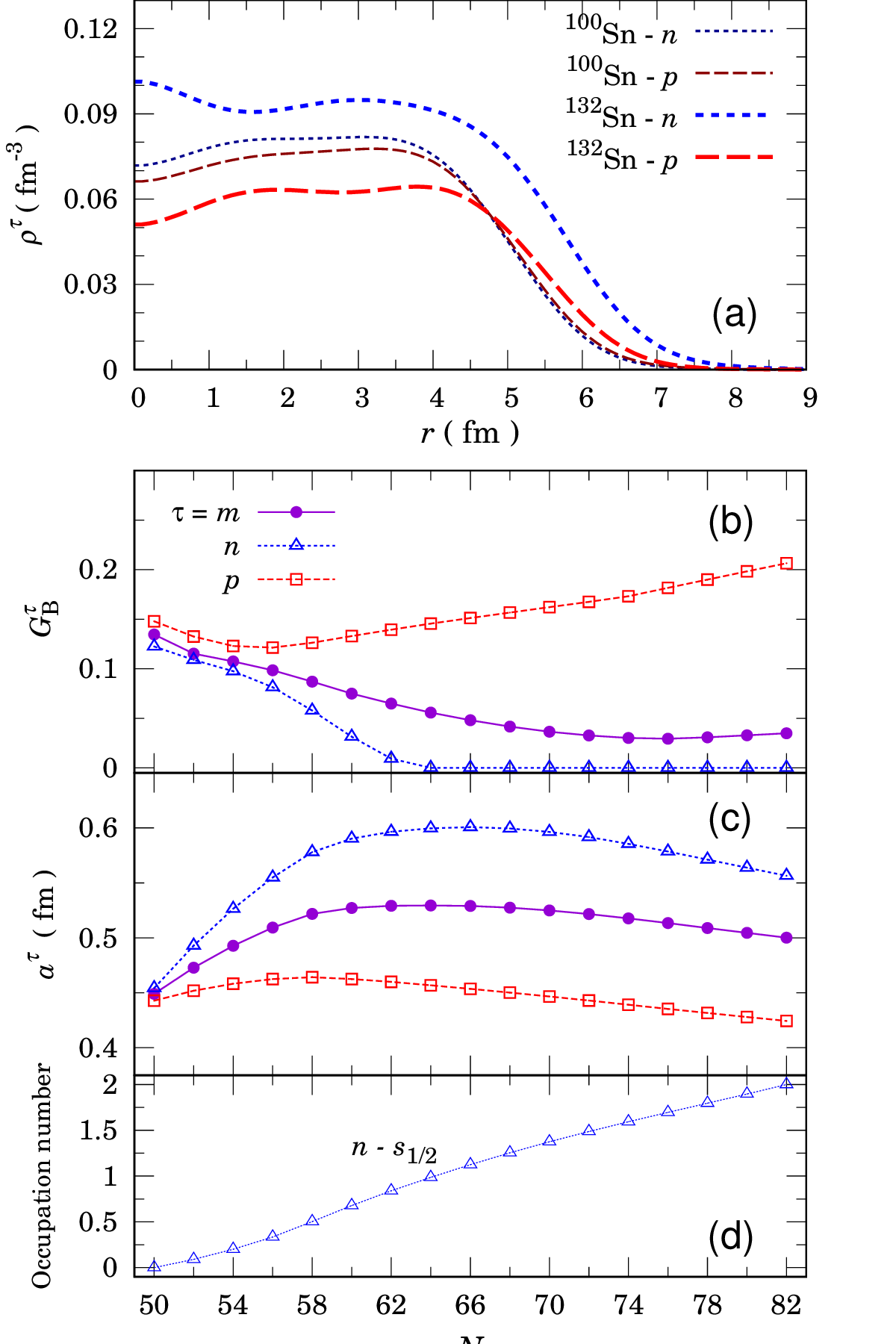}
\caption{Same as Fig.~\ref{BDnz14}, but for $Z=50$ isotopes. 
Panel (d) shows the occupation numbers of $3s$ state for neutron.
}
\label{DBDnz50}
\end{center}
\end{figure}
Next, we focus on the bubble structure of the medium-heavy nuclei [region (ii)], the $Z=50$ isotopes.
Figure~\ref{DBDnz50} plots the density profiles, bubble and diffuseness parameters, 
and occupation numbers of $s$-state for $N=50$--$82$ of Sn isotopes.
The neutron bubble parameter $G^n_{\rm B}$ is most pronounced at $N=50$ and decreases as increasing $N$. 
This can also be explained by the occupation of the $3s$ orbit 
because the excess neutrons fill the $s,d,g$ shell toward $N=82$. 
The evolution of the bubble parameter is smooth with respect to $N$, 
which is in contrast to that of light nuclei because the $s$ orbit is gradually occupied
with increasing $N$ due to the pairing correlation. 
This behavior of the bubble parameters is well-reflected in the nuclear diffuseness parameters. 
The diffuseness becomes large while $G^n_{\rm B}$ diminishes when the $s$ orbit is occupied.
In contrast, the proton bubble is enhanced as the neutron number increases
because the proton $2s$ orbit becomes more bound with the neutron excess.
The matter bubble behaves as an intermediate between the neutron and proton ones.

One may think that the bubble formation is related to the EoS parameter of nuclear matter.
To see the interaction dependence,
we investigate the correlation between the bubble and EoS parameters 
by employing eight standard Skyrme parameter sets selected by Ref.~\cite{Brown13}; 
SGII~\cite{VANGIAI1981379}, SkM$^\ast$~\cite{BARTEL198279}, 
SLy4~\cite{CHABANAT1998231}, KDE0v1~\cite{Agr05},
SkT3~\cite{TONDEUR1984297}, SV-sym32~\cite{Klu09}, 
LNS~\cite{Cao06}, and SkI3~\cite{Naz96}.
We take a covariance among typical EoS parameters and $G^{\tau}_{\rm B}$, 
but no robust correlation is found in this work.
This implies that the nuclear bubble structure is formed by the fluctuation of the shell structure 
rather than the bulk properties.
We remark that this finding is consistent with Ref.~\cite{Sch17}, in
which the central density lighter than $^{208}$Pb carries no information on the EoS parameters.

We, however, note that the nuclear bubble structure of $^{100}$Sn
is robust and emerges for the eight Skyrme parameter sets 
employed in this paper: well-developed bubble structure for proton,
neutron, and matter density distributions is predicted for $^{100-108}$Sn.
This matter bubble structure is reflected in the nuclear density profiles near the nuclear surface 
and can practically be extracted
by using proton-elastic scattering as prescribed in Refs.~\cite{PhysRevC.97.054607,PhysRevC.102.034619}.

\subsection{Bubble structure in the region (iii): rare earth elements}
The region (iii) corresponds to the rare earth elements region. 
Their bubble structure can basically be explained
by a combination of the mechanisms in regions (i) and (ii).
Many nuclei in the rare earth region have deformed shapes and the pairing correlation. 
Their character plays a role in modifying the occupation probabilities 
near the Fermi surface, which is reflected in the bubble parameters.

\subsection{Coulomb induced bubble structure in the region (iv): superheavy nuclei}
\label{sec:Cou}
\begin{figure}[h]
\begin{center}
\includegraphics[height=0.45\textwidth,keepaspectratio,angle=-90]{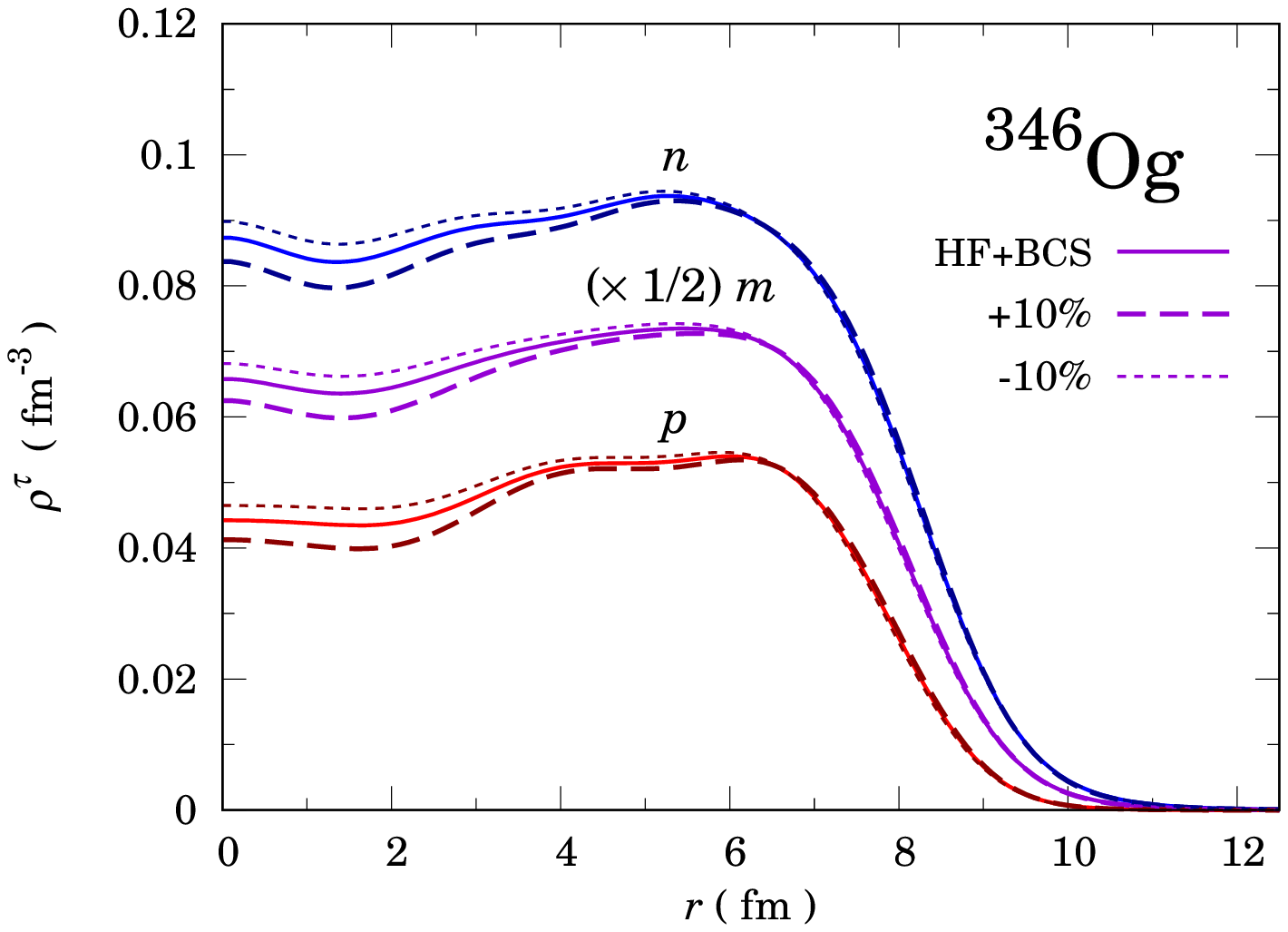}
\caption{Same as Fig.~\ref{DSn132}, but for $^{346}$Og.
 The results with varying the Coulomb strength by $\pm 10$\% are plotted for comparison.
}
\label{DOg346}
\end{center}
\end{figure}

Earlier study~\cite{Wil46} for the bubble structure of heavy or superheavy nuclei 
pointed out that the repulsive effect of the Coulomb force enhances the bubble structure.
We remark that, recently, this mechanism was microscopically explained by the wine-bottled proton mean-field 
induced by the strong Coulomb repulsion~\cite{PhysRevC.106.024321}.
To corroborate this, we examine changes in the internal density by artificially varying the Coulomb force.
Figure~\ref{DOg346} displays the density profiles of $^{346}$Og.
The density distributions obtained by changing the strength of 
the Coulomb interaction by $\pm 10$\% are also drawn for comparison.
As we see in the figure, the internal density is sensitive to the strength of the Coulomb force, 
while the radius at the maximum density is not for $^{346}$Og ($Z=118$) case. 
The $G^p_{\rm B}$ value is changed by $+0.064$ ($-0.032$), which corresponds to $+35$\% ($-18$\%),
for the Coulomb strength varied by $+10$\% ($-10$\%), respectively.
We also evaluated a case with smaller $Z$, $^{270}$Hs ($Z=108$).
The effect on the bubble parameter is less than the $^{346}$Og case:
The $G^p_{\rm B}$ value is varied by about $\pm 20$\% by the $\pm 10$\% changes of the Coulomb strength.
The strong Coulomb repulsion is essential to form the bubble structure in this superheavy mass region.

\subsection{Multi-layered bubble structure}
\label{sec:hollow}
Through the present analysis of the density profiles on the nuclear chart,
we find the higher order of the nuclear bubble structure, ``multi-layered'' bubble (MLB) structure, 
i.e., depressions of the density distributions other than the origin. 
The definition of the nuclear MLB is Eq. (\ref{eq:Gb}) in Sec.~\ref{sec:2.2}.

\begin{figure}[ht]
\begin{center}
\includegraphics[height=0.45\textwidth,keepaspectratio,angle=-90]{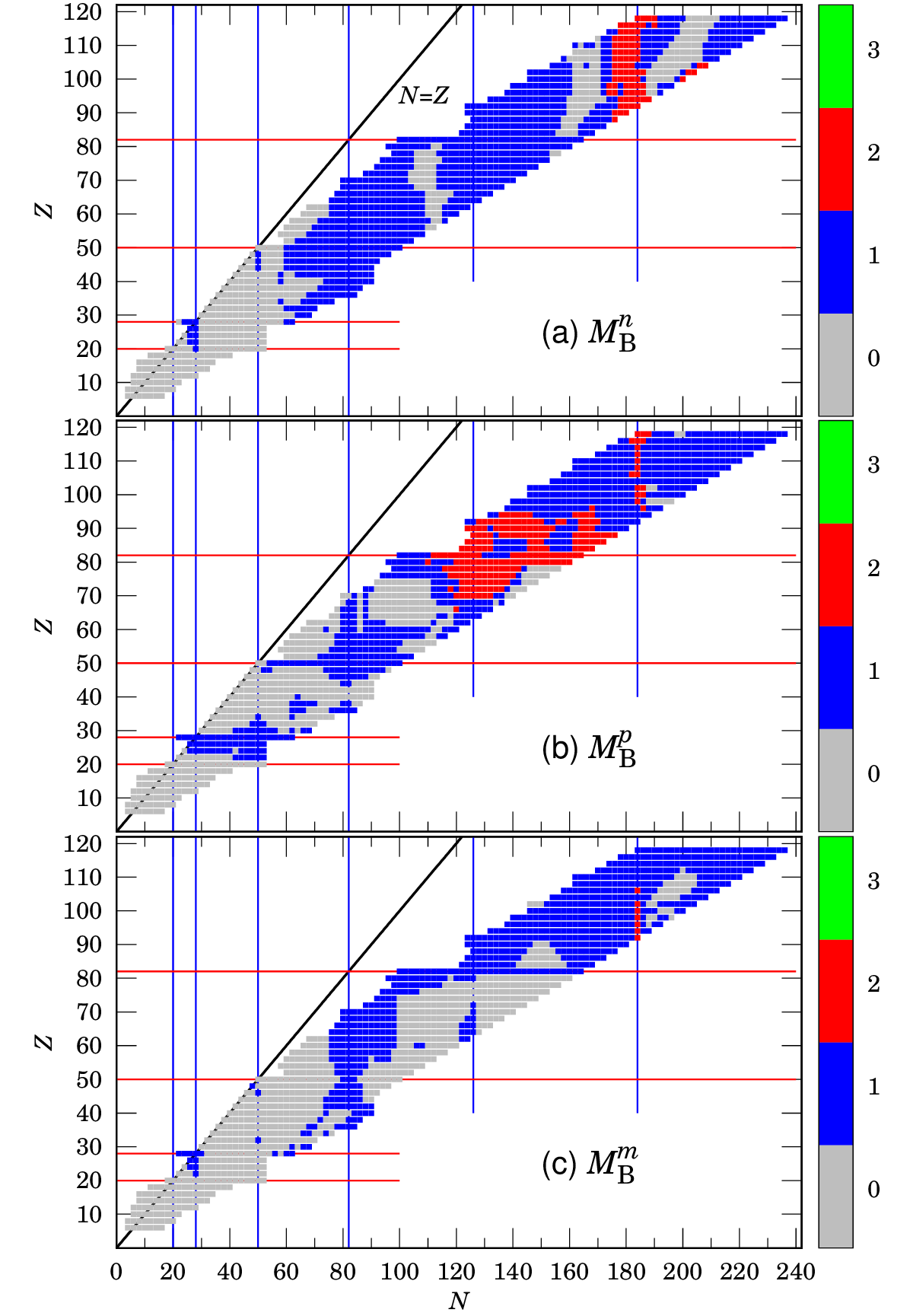}
\caption{Bubble-multiplicity $M_{\rm B}^\tau$ for (a) neutron, (b) proton, and (c) matter, 
excluding the bubble at $r_{\rm min}^\tau=0$.
}
\label{CavNanp}
\end{center}
\end{figure}

Figure~\ref{CavNanp} shows the number distribution of nuclear bubble-multiplicity 
on the nuclear chart for (a) neutron, (b) proton, and (c) matter,
in which the ordinary single-layered bubble structure is excluded.
Nuclei with more than three density depressions are not confirmed in this work.
The MLB structure appears at $A\gtrsim 40$ and
at around the magic numbers.
This implies that the shell structure also induces the emergence of the MLB structure as 
like the ordinary bubble structure.

\begin{figure}[h]
\begin{center}
\includegraphics[height=0.45\textwidth,keepaspectratio,angle=-90]{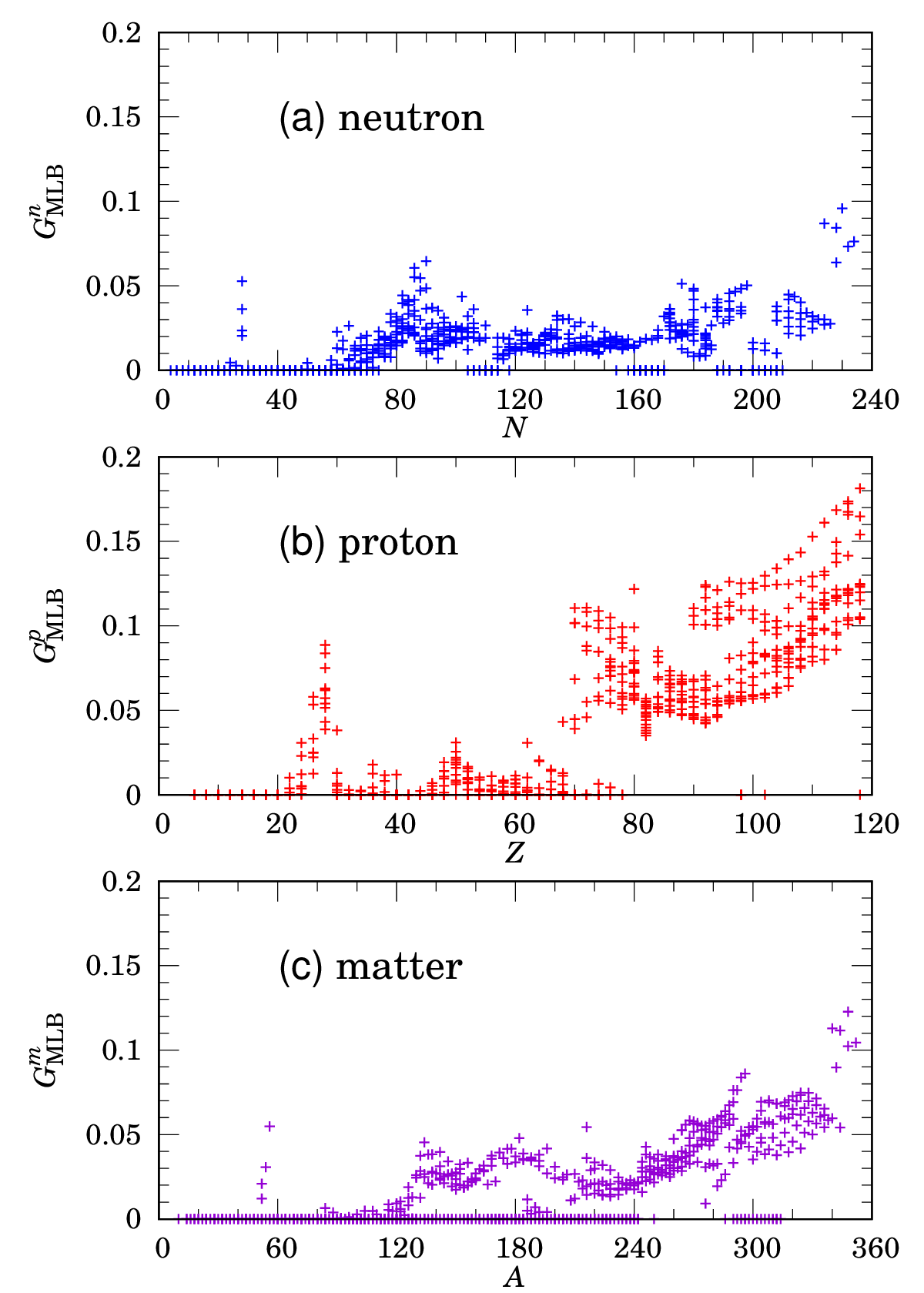}
\caption{Multi-bubble parameters $G_{\rm MLB}^\tau$ for (a) neutron, (b) proton, and (c) matter density distributions.
}
\label{CavPanp}
\end{center}
\end{figure}

Figure~\ref{CavPanp} shows the generalized bubble parameter $G_{\rm MLB}^\tau$ for (a) neutron, 
(b) proton and (c) matter density distributions as a function of the mass number. 
Although the values are, in general, smaller than the bubble parameters, 
the characteristic patterns appear for both neutrons and protons: 
$N,Z=28$, around $N,Z=50$, and $N,Z=82$, which correspond to 
$1f_{7/2}$, $1g_{9/2}$ and $1h_{11/2}$ closed configuration, respectively.

There is almost no MLB structure in the matter density distributions in $A\lesssim 100$. 
The nuclei having the neutron MLB structure are few in $N<50$,
and there is no nucleus having MLB structure for proton at $Z<20$.
The behavior of $G_{\rm MLB}^p$ differs from the others at $Z\gtrsim 82$.
We see some peak structures that may come from the shell structure at $Z\lesssim 60$, 
while it disappears in the heavy mass region $Z\gtrsim 82$ 
where the Coulomb force is expected to play a significant role.
For $Z\gtrsim 70$, the mechanism may also be attributed to the Coulomb force. 
Figure~\ref{CavPanp} shows a comparison between $G_{\rm MLB}^n$ and $G_{\rm MLB}^p$.
Actually, the $G_{\rm MLB}^p$ of the superheavy nucleus ($^{346}$Og) is
also changed by +0.085($-$0.038), which corresponds to $+44$\%($-19$\%) 
for the Coulomb strength varied by $+10$\%($-10$\%), respectively.
 
\begin{figure*}[ht]
\begin{center}
\includegraphics[height=0.8\textwidth,keepaspectratio,angle=-90]{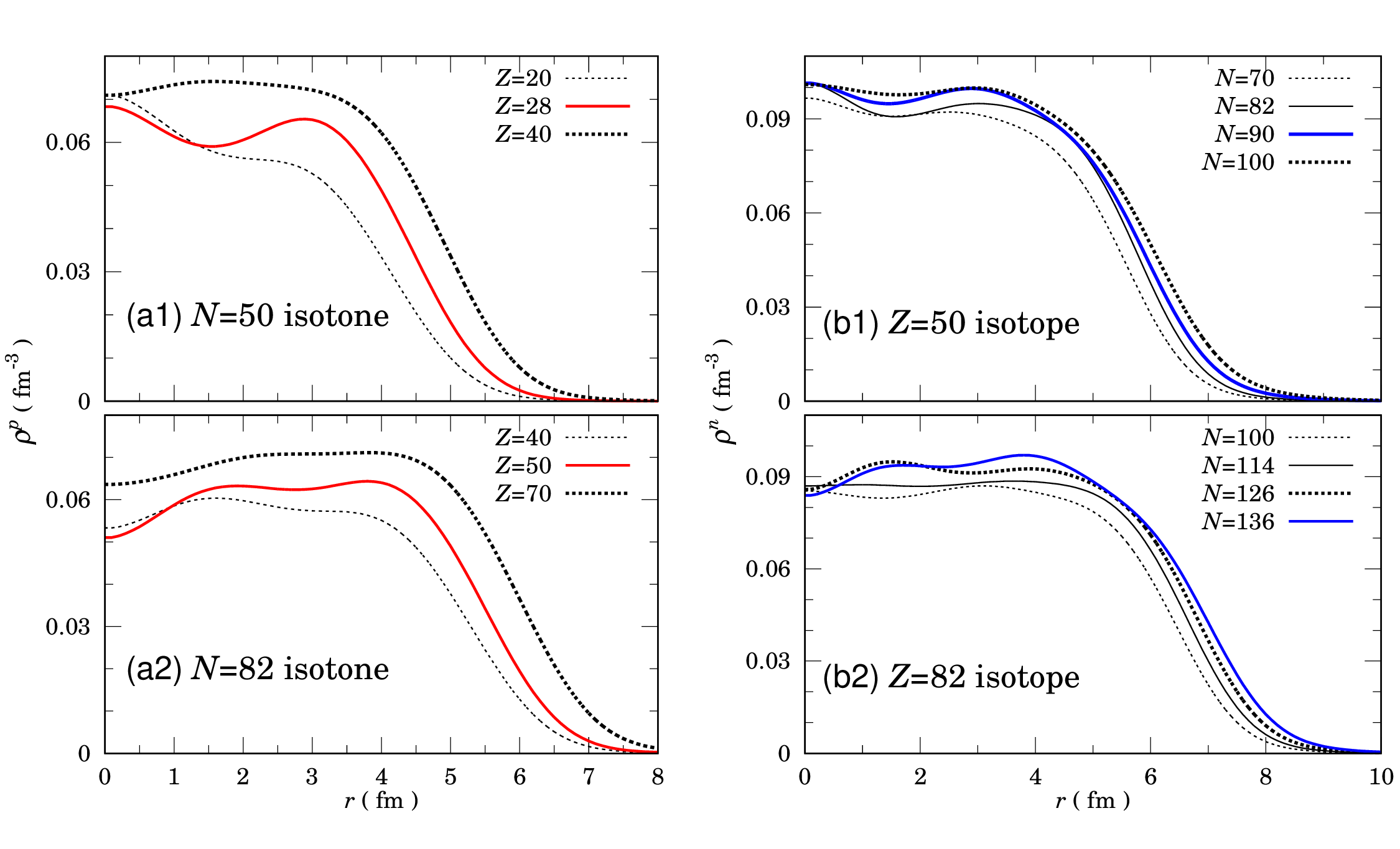}
\caption{
Density distributions of (a) $N=50,82$ isotones and (b) $Z=50,82$ isotopes.
}
\label{HolnpD}
\end{center}
\end{figure*}

Figure~\ref{HolnpD} plots the characteristic proton and neutron density distributions
for (a1) $N=50$, (a2) $N=82$ isotones and for (b1) $Z=50$, (b2) $Z=82$ isotopes, 
where that well-developed MLB structure is found.
We see some proton density depressions 
at 1.5 fm for $^{78}$Ni ($N=50$) in (a1) and at 2.8 fm for $^{132}$Sn in (a2), 
and some neutron density depressions at 1.6, 1.4 fm for $^{132,140}$Sn in (b1) and at 2.9, 2.4 fm for $^{208,218}$Pb. 

\begin{figure}[h]
\begin{center}
\includegraphics[height=0.5\textwidth,keepaspectratio,angle=-90]{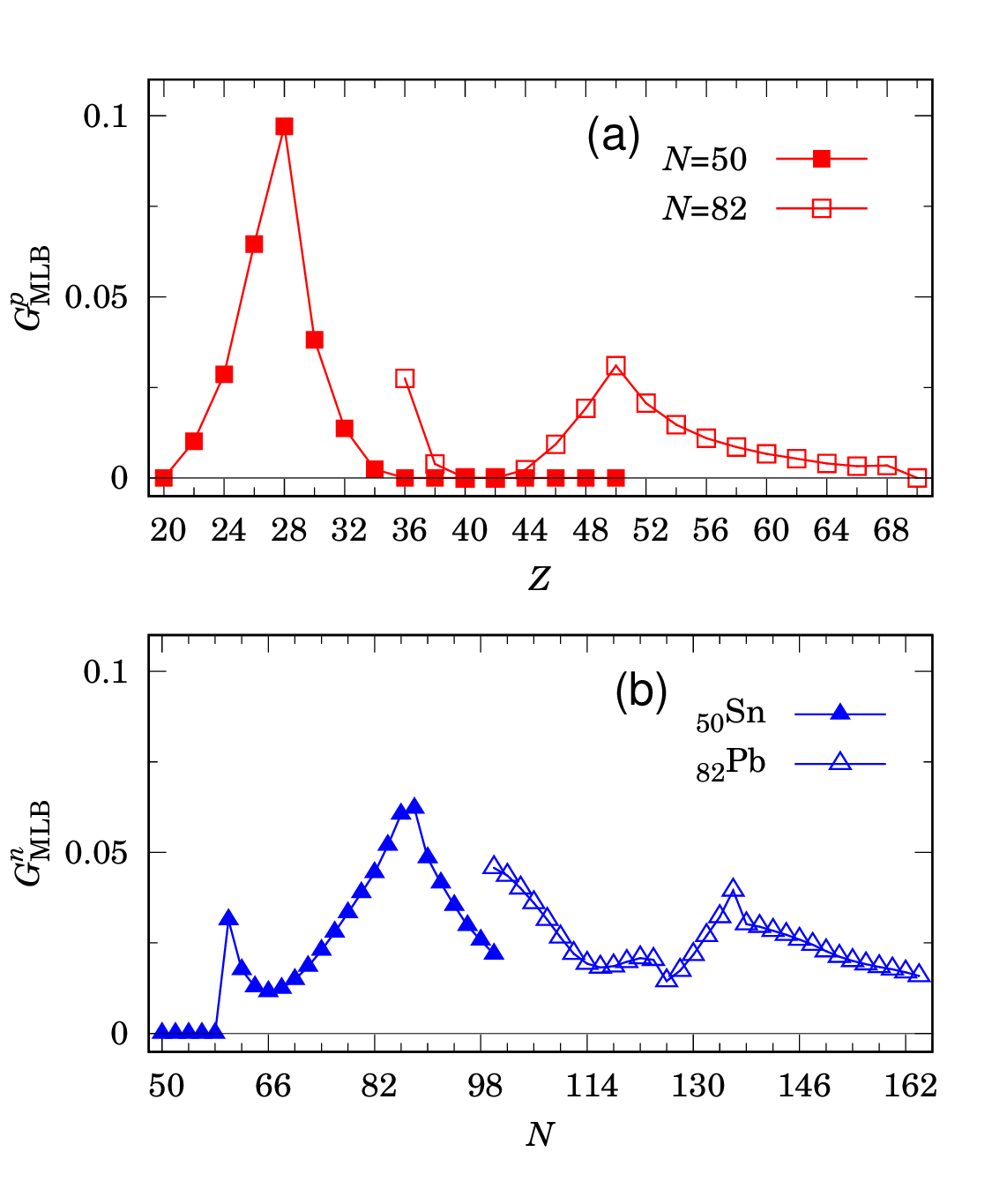}
\caption{
Nuclear MLB parameters $G_{\rm MLB}^\tau$ for (a) $N=50, 82$ isotones and (b) $Z=50,82$ isotopes.
}
\label{HolP}
\end{center}
\end{figure}

Figure~\ref{HolP} shows the MLB parameters (a) $G_{\rm MLB}^p$ for $N=50,82$ isotones 
and (b) $G_{\rm MLB}^n$ for $Z=50,82$ isotopes.
Here, we do not need to worry about the effects of nuclear deformation 
since all nuclei with $N,Z=50$ and $N,Z=82$ have spherical shapes
in the present model.
Let us first discuss the proton MLB structure.
In $N=50$ ($N=82$) isotopes, the $G_{\rm MLB}^p$ value is peaked at $Z=28$ ($Z=50$), 
where the $1f_{7/2}$ ($1g_{9/2}$) orbit is fully occupied.
This MLB structure rapidly disappears with increasing $Z$.
Since these single-particle orbits are nodeless $j$-upper orbits, 
their amplitudes are concentrated at the surface regions. 
In fact, the surface region of the proton density distribution with $Z=28$ 
is highly enhanced compared to that with $Z=20$, as displayed in Fig.~\ref{HolnpD} (a1). 
The proton density depression at 1.5 fm disappears for $Z=40$ as nodal single-particle orbits 
such as $2p$ orbits are occupied, which has some amplitude in the internal regions.
This contrasts the ordinary bubble structure,
in which the $s$ orbit plays a role in diminishing it.
  
The same behavior is also found for $N=82$ isotones,
though the degree of the MLB structure is not as prominent as that in $N=50$ isotones. 
The occupation of the $3s$, $2d$ proton orbits fills the density depressions around the origin and $\sim 3$ fm. 
The relatively larger $G_{\rm MLB}^p$ value at $Z=36$ is considered as
a remnant of the $Z=28$ MLB structure found in $N=50$ isotones in Fig.~\ref{HolP} (a).

For heavier nuclei, the emergent condition of the MLB structure becomes a little complicated 
because the pairing interaction plays a role.
For the $Z=50$ isotopes, the generalized bubble parameter becomes
large due to the occupation of the $1h_{11/2}$ states.
The $G_{\rm MLB}^n$ value is peaked at $N=90$ 
because the pairing correlations induce both $1h_{9/2}$ and $2f_{7/2}$ states,
while the $1h_{11/2}$ orbit is completely filled at $N=82$ if no pairing correlation is contributed.
For the $Z=82$ isotopes, the $G_{\rm MLB}^n$ value is not very large. 
The characteristics of the single-particle orbits are smeared out in such heavy nuclei.
The kink behavior at $N=126$ comes from the mixing of $1i_{11/2}$ and $2g_{9/2}$ orbits, 
which are essential to account for the sudden increase
of the charge radius as was discussed in Refs.~\cite{PhysRevC.91.021302,PhysRevC.105.044303}. 

\begin{figure}[h]
\begin{center}
\includegraphics[height=0.5\textwidth,keepaspectratio,angle=-90]{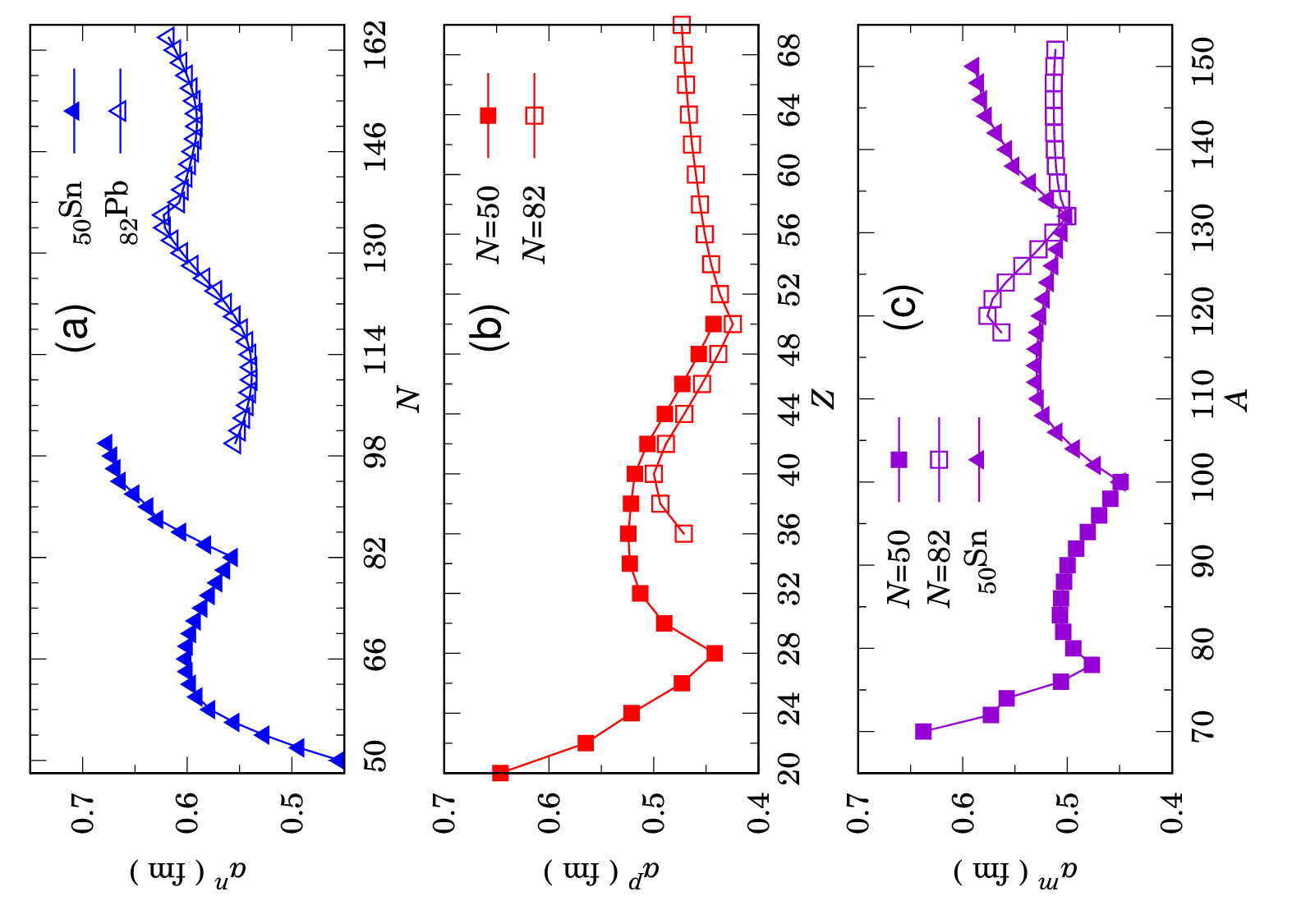}
\caption{Nuclear diffuseness parameters: 
  (a) neutron for $Z=50,82$ isotopes, (b) proton for $N=50,82$ isotones, 
  and (c) matter for $Z=50,82$ isotopes and $Z=50$ isotones.}
\label{DifP}
\end{center}
\end{figure}

Since the emergence of the MLB structure is closely related to the properties of the outermost single-particle orbits,
the generalized bubble parameter is well correlated with the diffuseness parameters.
Figure~\ref{DifP} shows the diffuseness parameters (a) for the neutron of $Z=50,82$ isotopes, 
(b) for the proton of $N=50,82$ isotones, and (c) for the matter of $N=50,82$ isotones 
and of $Z=50$ isotopes, respectively.
For the occupation of nodeless $j$-upper orbits, 
the nuclear surface becomes sharp as the angular momentum is the highest among the others, 
while the occupation of the nodal lower angular momentum single-particle states 
diffuses the nuclear surface~\cite{10.1093/ptep/ptab136}. 
In fact, as displayed in Fig.~\ref{DifP}, the peak position of $G_{\rm MLB}^{\tau}$ corresponds 
to the local minima of the diffuseness parameters.
A systematic measurement of the diffuseness parameters is useful to identify 
the MLB nucleus along the isotopic or isotonic chains.

We note that the MLB structure, including the ordinary bubble structure, 
is formed by the depletion of the specific single-particle orbits.
The deficiency of the $s$ orbit induces the nuclear bubble.
As a natural extension of this, the MLB structure is caused by the depletion of the nodal $p$ or $d$ orbits.
In fact, the degree of the MLB structure is enhanced for the nuclide 
with $Z=28$, $Z,N=50$, and $N=82$, which correspond to 
$2p$, $2d$, and $3p$-orbits depletion, respectively. 
Nuclei with $M_{\rm B}=2$ in Fig.~\ref{CavNanp} appear 
when both the $2p$ and $3d$ or both $3d$ and $4p$ orbits are depleted.

\section{Conclusion}
\label{sec:concl}
We have systematically investigated the non-uniformity of the internal density distributions of nuclei
for the whole nuclear mass region, 1,389 even-even nuclei with from $Z=6$ to 118,
using the HF+BCS model represented in the 3D coordinate space. 
The nuclear deformation and pairing correlation are taken into account.
To quantify the depressions of the nuclear density distributions,
we introduce the generalized bubble parameter
to evaluate both the ordinary bubble structure
and the multi-layered bubble (MLB) structure, which exhibits
some depressions in the density distributions at $r\neq 0$.

We discussed the bubble structure categorized into the following four mass regions:
(i) at $N=14,16$ and $Z=14,16$, (ii) at around Sn isotopes,
(iii) $Z<82, N>100$, and (iv) superheavy elements. 
The emergence is correlated with the vacancy of the $s$ orbits,
which is strongly affected by nuclear deformation and pairing correlations.

We show that in light nuclei, e.g., $N,Z=14,16$,
the nuclear shape plays a decisive role in determining
the occupation of the $s$ orbit, leading to the sudden emergence
of the bubble structure. This spectroscopic
information is reflected in 
the density profiles near the nuclear surface.

For $N,Z=50$ nuclei, the pairing correlation plays a role and induces 
the gradual evolution of the nuclear bubble structure.
We find that its pronounced bubble structure of $^{100}$Sn, 
which is robust, as it appears, with all the Skyrme density functionals employed in this paper.
   
For the region (iii), the emergent mechanism can be explained 
by a combination of the ones for the regions (i) and (ii). 
For the superheavy elements ($Z>100$), we show that the strong Coulomb force assists 
the non-uniformity of the density distributions.

As an extension of the bubble structure, the MLB structure is investigated systematically
on the nuclear chart and is found in a wide nuclear mass region.
The proton density distribution of $^{78}$Ni shows a typical MLB structure, having a local minimum 
at $r_{\rm min}^p\neq 0$ in the density distributions.
We show that the emergence of the MLB structure is correlated with the occupation of nodeless $j$-upper orbits 
because the nodeless $j$-upper orbits are concentrated at the surface regions that 
can attract the nucleons in the internal regions.

The nuclear non-uniformity appears due to the lack of specific single-particle orbits, 
which are the nodal $s$, $p$, and $d$ orbits.
Therefore, the degree of the non-uniformity may depend on a structure model employed, 
since each model has the difference of the single-particle states.
Nevertheless, some nuclei exhibit robust bubble or MLB structure, 
which does not depend on the effective interaction employed in this paper.
Determining the spectroscopic information is
interesting to quantify the degree of the non-uniformity
of the nuclear density distributions.
Even if such a direct measurement is difficult,
the non-uniform internal density is reflected in the density profiles 
near the nuclear surface, which can be observed
by proton-elastic scattering measurement~\cite{PhysRevC.97.054607}.
  
\acknowledgments
This work was in part supported by MEXT Leading Initiative
for Excellent Young Researchers Grant, Japan, and
JSPS KAKENHI Grant No.\ 20K03943, 23K22485, and JSPS Bilateral Program Number JPJSBP120247715.

\bibliography{reference}
\end{document}